\documentclass[12pt,preprint]{aastex}
 
\def\a{$\alpha$}

\def\m{$\mu$m}
\def\ab{$\sim$}
\def\p{$\pm$}
\def\dla{damped Lyman $\alpha$}
\def\Dla{Damped Lyman $\alpha$}

\begin{document}
\title{NICMOS Snapshot Survey of \Dla \ Quasars}
 
\author{James W. Colbert, Matthew A. Malkan}
\affil{UCLA Dept. of Physics \& Astronomy}
\affil{University of California, Los Angeles, CA 90095}
\email{colbert@astro.ucla.edu, malkan@astro.ucla.edu}

\begin{abstract}

We image 19 quasars with 22 \dla \ (DLA) systems using the F160W filter and the Near-Infrared Camera and Multiobject Spectrograph aboard the {\it Hubble Space Telescope}, in both direct and coronagraphic modes. We reach 5$\sigma$ detection limits of \ab H=22 in the majority of our images. We compare our observations to the observed Lyman-break population of high-redshift galaxies, as well as Bruzual \& Charlot evolutionary models of present-day galaxies redshifted to the distances of the absorption systems. We predict H magnitudes for our  DLAs, assuming they are producing stars like an L$_{*}$ Lyman-break galaxy (LBG) at their redshift. Comparing these predictions to our sensitivity, we find that we should be able to detect a galaxy around 0.5-1.0 L$_{*}$(LBG) for most of our observations. We find only one new possible candidate, that near LBQS0010-0012. This scarcity of candidates leads us to the conclusion that most \dla \ systems are not drawn from a normal LBG luminosity function nor a local galaxy luminosity function placed at these high redshifts.    

\end{abstract}


\section{Introduction}

In recent years, our knowledge of the high redshift universe has finally grown to include a significant number of galaxies powered mainly by star formation, often referred to as Lyman-break galaxies because most were discovered by searching for the Lyman-limit discontinuity using broad band colors \citep{ste96}. There is still the concern, however, that these galaxies may not be representative of the typical galactic mass at that epoch. Dust, for example, could strongly affect the discovery rate of galaxies found using searches, like the Lyman-break method, that search in rest frame ultraviolet light \citep{row97}. 

Quasar absorption line studies, on the other hand, have the advantage of being able to follow the neutral hydrogen content of the universe regardless of whether it emits light or not. While absorption line studies may also suffer from selection effects, with dustier intervening gas creating dimmer quasars that are either less well studied or not discovered at all \citep{pei95, car98}, they represent a potentially less biased tracer of the content of the universe than the Lyman-break galaxy. Examinations of \dla \ (DLA) systems, which make up the bulk of all absorption line neutral hydrogen, show that at high redshifts, the mass of neutral hydrogen seen in DLAs per unit comoving volume is roughly the same as the mass density seen in luminous matter (i.e. stars) in present-day spiral galaxies \citep{lan91}. Since this is a much larger quantity of neutral gas than we observe in galaxies today \citep{rao93, zwa98}, this suggests that much of the gas we see in high z DLAs has ended up in stars. In fact, \cite{wol95} identify the DLA systems as the likely progenitors of current spiral galaxies. 

However, some recent work has challenged this idea, showing no drop in the cosmological mass density of neutral hydrogen from redshifts of 3 down to 0.5 \citep{rao00}, instead of the expected gradual decline seen in other work \citep{smi96,jan98,sto00}. Another indication that DLAs may not be forming today's galaxies is the lack of any metal enrichment in lower redshift DLAs, with the mean Zn/H ratio remaining around 1/10 solar from z=1-3 \citep{pet99}. It should be noted that the statistical uncertainties involved in these measurements remain significant, since all work on DLAs at low redshifts relies on a small number of objects (23 with z$<$1.65 for \cite{rao00}), with the handful with highest column densities dominating the statistics. These low redshift DLA studies are also the most vulnerable to bias from any possible obscuration of quasars by dust in foreground absorbers \citep{sto00, pei95}.    

Modeling of velocity profiles has not settled the question. Supporting the contention that DLAs form spiral galaxies, \cite{pro97} demonstrate how models of rotating disks, not unlike today's spirals, fit the observed double trough velocity profiles seen in some DLA metal lines. However, there are many other viable models that also produce similar velocity profiles, such as merging protogalactic clumps \citep{hae98}, randomly moving clouds in spherical halos \citep{mcd99}, and multiple gaseous discs in a single halo \citep{mal00}. 

Another problem with the DLAs as spiral galaxies hypothesis is the number of DLA systems that are being discovered, compared with the number of spiral galaxies around at the present day. Based on the present day density and size of spiral galaxies, one can estimate the number that should intercept a random line of sight to a quasar at some high redshift. While the exact prediction depends on the assumed cosmology, \cite{lan91} found that the number of DLA systems being discovered is at least twice that predicted, even under the more favorable cosmological assumptions. This means that if DLAs are high-redshift spiral galaxies, then there must be strong evolution in the spiral population with redshift. Either spiral galaxies were more numerous in the past, or they were larger or both. Another way around this discrepancy is to assume that our understanding of local galaxies and the measured fiducial parameters used for these predictions are flawed. For instance, a large population of low surface brightness galaxies \citep{imp89} or low luminosity dwarf galaxies \citep{cen98} could dominate the absorption cross sections responsible for \dla \ absorption lines. This, however, is not the case for low-redshift Mg II absorption line systems for which visible counterparts have been found, where the majority are essentially bright galaxies with typical luminosities just under L$_{*}$ \citep{ber91}. 
      
It would be valuable to find some connection between DLAs and high-redshift galaxies, our two largest sources of information on the non-AGN high-redshift universe. Unfortunately, while we know a great deal about the distribution, metallicity, and evolution of the neutral gas in high-redshift DLAs, we know almost nothing about their associated stars, galactic sizes, or morphologies. Only by imaging starlight from these DLA systems can we measure how much, if any, star formation is actually occurring in these objects. Lyman break galaxies produce stars at a steady pace of about 10 M$_{\sun }$ yr$^{-1}$, depending heavily on cosmology and the large dust extinctions assumed \citep{see96}, not that different from the \ab 10 M$_{\sun }$ yr$^{-1}$ seen in H$\alpha$ emission line galaxies today \citep{gal95}. If DLAs are progenitors of present day galaxies and/or the same population as the Lyman break galaxies, then they should also be producing stars and starlight. This is true whether the DLAs are galactic disks, spheroids, or some other assembled structure.

Searches for visible candidates for low to medium-redshift DLA systems have had mixed results. Often candidates are found, but less than half of them are suspected to be reasonably bright spiral galaxies. The rest are a non-uniform collection of low surface brightness galaxies, bright compact objects, and dim dwarfs \citep{ste94,leb97,lan97,rao98,pet00}. Again the statistics are poor (less than 15 objects) and most have not been spectroscopically confirmed, leaving the nature of these DLAs still murky. 

High-redshift systems have been even more problematic, with much lower candidate discovery rates. A major reason is that this is a difficult measurement, considering the large distances to these objects and their close projected proximity to a bright quasar. A common search technique has been to look for emission lines, Lyman \a \ or H\a , where the emission line flux ought to stand out from the background and the quasar light may be depressed. These searches have found at least two fairly unambiguous detections, near quasars PKS0528-250 \citep{mol93} and 2231+131 \citep{djo96}, but broader searches of DLA samples have generally found nothing at all \citep{low95,bun99}. Another strategy for imaging DLAs has been to go into the infrared, where K-corrections would be more favorable and dust, often cited as a possible explanation for the failure of Lyman \a \ searches, would likely be only a minor effect. For instance, \cite{ara96} observed ten quasars with DLA systems in the near-infrared, discovering two candidates (in front of quasars 0841+129 and 1215+333) after subtracting off the PSF of quasar. Since both candidates lie about an arcsecond away from the quasar, the PSF subtraction is critical and an error could lead to a false detection. Higher resolution imaging would greatly help. 

We have obtained imaging of 19 quasars with confirmed DLA systems using the Near Infrared Camera and Multi-Object Spectrometer (NICMOS) onboard the Hubble Space Telescope (HST). This gives us the advantage of both the near-infrared, with its favorable K-corrections and low dust dependence, and high spatial resolution, which will separate any candidates from the quasar's light, even at distances of less than arcsecond. In addition, the majority of the observations were done using the NICMOS coronagraph, which greatly decreases scattered light from the quasar, improving chances of candidate detection. If DLAs are high-redshift disk galaxies producing stars, missed only because of their close proximity to the quasar, a survey of this type can uncover them. Another advantage of a large survey of a sample like ours is that it presents the failures as well as the successes, something that is often missing from papers on individual candidate discoveries. A rate of DLA system discoveries gives us a statistically significant result for the entire DLA class of objects, something an individual candidate can not. Section 2 describes our sample and the observations, while $\S$3 describes the data reduction and analysis. Section 4 discusses the sensitivity of our survey to these hypothetical L$_{*}$(LBG) galaxies. Section 5 presents the results of our candidate search. Finally, $\S$ 6 presents our discussion and conclusions. (A list of notes on individual DLAs is provided in the Appendix).

\section{Sample, Observations, \& Data Reduction}

\subsection{Sample Selection}

Our target candidates were selected from all confirmed DLAs with n(HI) $>$ 2 $\times$ 10$^{20}$ cm$^{-2}$ not already being imaged in other HST programs. These include surveys by \cite{wol86}, \cite{tur89}, \cite{lan91}, \cite{lu94}, \cite{wol95}, \cite{sim96}, and \cite{sto96}. The resulting sample contains DLAs with redshifts ranging from 0.86 to 3.7, with a median z = 2.10,  and column densities of hydrogen in the range 20.2 $<$ log n(HI) $<$ 21.8, with a median of 5 $\times$ 10$^{20}$ cm$^{-2}$. The 19 DLA quasars observed were selected to fill any gaps in the HST NICMOS observing schedule, with no weight or preference given to any particular object. This sample includes three quasars (LBQS2206-1958, $[$HB89$]$2348-011, \& LBQS2359-0216B) with two DLAs each. Also included in the 19 sample DLA quasars is a single low redshift \dla \ object, $[$HB89$]$0809+483, with a absorbing z of 0.437. A couple of possible optical counterparts have been well established for this object (R\ab 20.3 \& R\ab 22.6; \cite{leb97,coh96}), making it valuable as a check that our methods can at least find known galaxies. Indeed they are easily seen, having a near-infrared magnitudes of H=17.5 and 19.6. All 22 target DLAs are listed in Table 1. 

Snapshot observations began March 16, 1998 and continued until November 1998.
   

\subsection{Coronagraph Observations and Data Reduction}

The majority of snapshots were taken using the NICMOS coronagraph, placing the quasar beneath the coronagraphic hole to reduce its scattered light. All observations were made using the NICMOS camera 2, which has a 19.2\arcsec by 19.2\arcsec \ field and a pixel scale of 0.075\arcsec , and the F160W filter ($\lambda _{central}$ = 1.594 \m , $\Delta \lambda$ = 0.403 \m ), which is roughly equivalent to the {\it H} band. The placement of the quasar was further complicated by the migration of the coronagraphic hole, a result of the NICMOS dewar anomaly \citep{sch98}. To properly place the quasar beneath the coronagraphic hole, an 110 second acquisition image had to be taken first. The NICMOS acquisition software must find both the hole and the target quasar for each observation. In several cases where the quasar was much dimmer than predicted, the software misidentified a high background spike as the target quasar and attempted to move that beneath the quasar hole. Unfortunately, in these cases it usually moved the target quasar entirely off the image, removing our ability to get even a direct image of it.

All observations used the MULTIACCUM STEP256 sampling sequence (See NICMOS Handbook, \cite{mac97}). For the coronagraph images, we took one exposure using NSAMP=13 for a total of 768 seconds. The MULTIACCUM data sets were reduced using the NICREDUCE software \citep{mcl97} and the flat-field, mask, and dark frame reference files provided by Space Telescope Science Institute. Data reduced in this way was superior to the standard CALNICA pipeline \citep{voi97}, with better cosmic ray subtraction and fewer bias/flat-field artifacts. Even after this reduction, most images still had a remaining ``pedestal effect'' or small residual DC offset not removed by dark current subtraction that shows up as an imprint of the flat-field. It is different for each the NICMOS detector's four quadrants. To correct for this remaining effect we used the Pedestal Estimation and Quadrant Equalization Software developed by R. van der Marel. This program takes the calibrated data and its flat-field and attempts to minimize the spread in values for pixels by assuming different values for an uncorrected, constant bias in each quadrant separately. 

There was usually a significant amount of scattered light left in the coronagraph images out to 2-3\arcsec , making PSF subtraction necessary. Here we used PSFs from the SMOV 7052 NICMOS coronagraphic performance verification run (P.I., G. Schneider), where they placed a bright non-binary stellar target (BD+032964, H=5.0) behind the coronagraphic hole. We scaled this coronagraph PSF to each quasar's magnitude and subtracted it, leaving only minor residuals. Examples of this light subtraction technique is shown Figure 1. For the brightest objects the subtraction did leave behind a couple of artifacts, which we identified by their repeated appearance in the same location in multiple quasar images. These artifacts all lay right at or just beyond the nonusable limit of the coronagraphic hole (radius \ab 0.4\arcsec ), due to inequalities along edge of hole. Only three observations (JVAS1757+7539, $[$HB89$]$1157+014, \& $[$HB89$]$0454+039) had such significant residuals left that they strongly affected object identification that close to the quasar. 

\subsection{Direct Observations and Data Reduction}

In three situations, where there was a strong possibility of capturing a nearby star in the same frame, we opted to make direct observations instead of placing the star under the coronagraph. The nearby star captured at the same time provided an excellent simultaneous PSF for subtraction from the quasar, allowing us to see even closer inwards than the coronagraphic disk allowed. A PSF taken at the same time as the image is valuable, as the detailed character of the NICMOS PSF changes with time \citep{kri98}. Additionally, utilizing more than one method assured us some data, in case one method failed. These observations also used F160W and the NICMOS camera 2.

Direct imaging consisted of two exposures, each using MULTIACCUM STEP256 with NSAMP=12, for a total of 1024 seconds. We dithered the two images 0.4\arcsec (5 pixels) from each other in order to help remove bad pixels and cosmic rays. Larger dithering patterns would have endangered our ability to keep both quasar and star within the image frame. 

Direct images were reduced in the same way as the coronagraphic observations, except for the PSF subtraction. For direct images the first choice for the PSF was any star available on the same frame, scaled to the magnitude of the quasar. In the case of [HB89]1209+093, no such star was available, so we used PSFs from the other observations where we did observe stars. In all cases we tried many different PSFs, using both stars and quasars, different magnitude scalings, and slightly different alignments in order to maximize the quality of the PSF subtraction. An example of this light subtraction technique is shown in Figures 2.
 
\section{Data Analysis}

We ran SExtractor \citep{ber96} on the final reduced and quasar PSF-subtracted images, detecting anything with four connected pixels each 1.5$\sigma $ above the noise. We used a 0.95\arcsec \ diameter aperture, roughly twice the median FWHM of all the discovered galaxies, for the photometry. We then made direct measurements of the standard deviation of groups of four pixels for all four quandrants. Some observations had noisier sections on their images, particularly in the lower left and along the bottom of the image (\ab 40\% higher). Less frequently there was an area of higher noise around the coronagraphic hole as well (\ab 10-33\% higher). To be conservative, we picked the highest $\sigma$ measurement, about 0.4 magnitudes above the minimum. Calculating the magnitude for a 5$\sigma $ detection for each section of each image, we went back through and cut all detections that did not meet that criteria. The typical 5$\sigma $ detection limit within our aperture was \ab H=22.1. We considered using a 3$\sigma $ cut-off instead, but found that the NICMOS images had too many tiny noise artifacts to trust such borderline detections. To finish, we made a final pass through the images, removing all obvious unreal detections (long, thin lines, impossibly small FWHMs, hot pixels, etc.), borderline objects on the image edge, and known coronagraph artifacts. 

The magnitudes listed for all detections in Table 2 have been changed from magnitudes within our 0.95$\arcsec$  diameter aperture to total magnitudes, using small corrections determined by placing sample galaxies at random throughout our images and then retrieving them with SExtractor. These sample galaxies resemble the galaxies detected, but with different FWHMs and known total magnitudes, allowing us to determine a relation between measured magnitude, FWHM, and the final corrected or true magnitude. This also allowed us to determine the limit to which galaxies can be reliably extracted, which is indeed around H\ab 22, depending on the FWHM of the sample galaxy. The detection limits listed in Table 1 have been similarly changed from their 0.95$\arcsec$ diameter aperture limits assuming a 0.4$\arcsec$ FWHM, the median FWHM of all our dim (H$>$20) galaxy detections, giving a typical 5$\sigma$ detection limit of H=22.0. Figure 3 displays PSF-subtracted close-ups (roughly 5$\arcsec$$\times$5$\arcsec$) of the area around the quasar for all successful coronagraph observations, while Figure 4 does the same for the direct observations.

\section{Sensitivity}

We now determine how bright we expect these DLA galaxies to be, and whether our survey is sensitive enough to see them. The best approach would be to compare our sensitivities with measured H magnitudes of star forming galaxies already found at high redshift. Unfortunately, there are very few spectroscopically confirmed star-forming galaxies at z=2.1, the median redshift of our DLA systems. Instead we will compare our objects to the large body of data on Lyman break galaxies (LBGs) at z\ab 3, the main source of information for ``typical'' galaxies in the high-redshift universe. Lyman break galaxies typically produce stars at a higher rate (\ab 10 M$_{\sun }$ yr$^{-1}$; \cite{see96}) than present day L$_{*}$ galaxies (\ab 1 M$_{\sun }$ yr$^{-1}$; \cite{ken94}). However, we expect most galaxies to undergo much greater star formation in their earlier, formative stages, making LBGs our best available candidates for modern galaxy progenitors. Alternatively, if we compare our DLAs to known galaxies at low redshift, where we also have a lot of data, we will need to make more critical assumptions about galaxy evolution.

\cite{ste99} provide a list of 564 spectroscopically confirmed Lyman break galaxies, with a median redshift of 3.04$\pm$0.24. Fitting a Schechter function to their data, they found their L$_{*}$ magnitude for LBGs to be ${\cal R}$=24.48$\pm$0.15, where ${\cal R}$ is the filter described in \cite{ste93} and given in AB magnitudes. To convert from ${\cal R}$ to the more standard R for faint galaxies, \cite{ste93} give the simple relation ${\cal R}$=$R$+0.14. $R_{AB}$ is almost identical to $R_{Vega}$ (only -0.055 mags difference), so our L$_{*}$(LBG) galaxy magnitude is 24.28 in $R_{Vega}$. The limited near-infrared observations of LBGs suggest a typical $R-H$ color \ab 2.2 \citep{dic00}, or an expected near-infrared magnitude for an L$_{*}$(LBG) z=3 galaxy of \ab22.1. This makes it unlikely that we can discover an emitting galaxy for our most distant DLA objects (2 are greater than z=3), since our typical 5$\sigma$ limit is around $H$=22, unless the DLA galaxy is bright. The majority of our sample, however, is at lower redshifts, so the real question is what magnitude we would expect if a typical z=3.04 LBG were found at our median z=2.1.
 
The expected distance modulus change from $z=3.04-2.1$ varies depending on cosmology selected. For this paper we are assuming a q$_o$=0.1, H$_o$=70 km s$^{-1}$ Mpc universe for which the distance modulus difference is 1.11 magnitudes. To account for the change in observed rest frame from 4000 to 5000\AA , we must apply K-corrections that depend on the type of galaxy one assumes. For the H-band between $z=2$ and 3, the K-correction is smaller for later type galaxies, becoming larger as the stellar population gets older and redder. We will assume young star-forming Sb-like galaxies, both because that should be appropriate for Lyman break galaxies and because it is the conservative assumption. The K-correction for an Sb galaxy between a $z=2.1$ and 3 is \ab 0.6 magnitudes \citep{pog97}, making the galaxy brighter. Combining both the distance modulus and K-correction gives us a predicted magnitude for an L$_{*}$(LBG) galaxy of H$\simeq$20.4. So, if we place the known distribution of $z=3$ LBG galaxies at our median redshift, we should have no problem detecting it.

This is not a realistic scenario, however, since one would certainly expect some sort of evolution in the spectral energy distribution (SED) of the LBG galaxies from $z=3$ to 2. An increase in star formation will brighten the galaxy, while a decrease or cessation of star formation will dim it. H band, which observes rest-frame optical, is less effected by evolution than the R band, which measures the rest-frame ultraviolet. However, even in the near-infrared, a passively evolving galaxy can be magnitudes different from a recent starburst. Since we have little information on the star formation or colors for galaxies over most of the redshift range these DLAs lie (z=1-2.5), we need to rely on models that at least reproduce the average colors and star formation we see in galaxies today. For this we use the evolutionary model SEDs of \cite{bru96}, redshifted to the predicted distances of the DLA objects and then convolved with the transmission curve of the H filter to produce our predicted magnitudes. These models use the stellar atlases of \cite{gun83} compiled by \cite{bru93}, a Salpeter initial mass function, and solar metallicity. 

We use an exponentially decreasing star formation rate (SFR) to simulate the star formation history of the possible galaxies. \cite{bru93} demonstrated that such exponentially decreasing SFRs reproduce present day galaxy spectra well, with exponential timescales increasing from around 1 to 7 Gyr going from E to Sc galaxies. We choose a timescale of 4 Gyr, appropriate for an Sb galaxy. This is not a unique solution, however, as combinations of single bursts and/or constant star formation can also produce present day galaxy SEDs. We assume our star formation started at z=5, but it should be noted that not much time passes between a z=5 and z=10 and therefore exactly when evolution starts makes little difference in the model results (about 0.1 mag in R-H, for example) as long as star formation is well under way before the range of z's we are probing (z=1-3.5). Again, this is a conservative assumption, for galaxies only become redder if formed at higher redshifts. 

Finally, we require that our models fit the observed colors and magnitudes observed for LBGs at z=3.04. \cite{ste99} has already demonstrated this is possible, fitting their colors with Bruzual and Charlot models of continunous star formation, reddened by dust with a corresponding $E(B-V)$=0.15. We adopt the same dust reddening for all our DLA objects, although we use the dust extinction formulation of \cite{car89} for our galaxy, not the starburst law of \cite{cal97} used by \cite{ste99}. However, except for normalization, these two laws produce essentially the same answers for the optical through infrared rest wavelengths examined by our 1.6 $\mu$m filter. Even though the DLAs span a large amount of cosmic time, we justify the use of a constant amount of reddening throughout because DLAs show very little, if any, metallicity evolution over the same redshifts \citep{pet99}. This is also a conservative assumption, as any increase in dust reddening would lead to larger R-H colors, making the infrared a more sensitive probe of high redshift star formation. 

As an alternative hypthesis, we also investigated taking an L$_{*}$ galaxy today and evolving it back to the redshifts of our DLAs. To determine present day L$_{*}$, we use the R band measurements of \cite{lin96} and a R-H color of 2.4 \citep{man01}. This gives a $z=0.02$ magnitude of H=11.2 (H$_o$=70). We then evolve this L$_{*}$ galaxy backwards, using the same exponential \cite{bru96} models and dust extinction that we used to evolve the LBGs forwards in time. These determinations of L$_{*}$ evolved back in time give almost the same results as L$_{*}$(LBG) brought forward, predicting galaxies only 0.2 magnitudes brighter. Evolving forward from LBGs at $z=3$ is therefore slightly more conservative and we will use L$_{*}$(LBG) for all comparisons. These final predicted L$_{*}$(LBG) H magnitudes are listed in Table 1.

We have plotted the predicted L$_{*}$(LBG) magnitudes for all our observed DLA objects versus the 5-$\sigma$ magnitude limits for each of the observations in Figure 5, excluding only the z=0.437 DLA object from $[$HB89$]$0809+4822, which would lie far off the left side of the graph because of its relatively low redshift. This plot shows what galaxies each observation would detect. The solid line represents where the magnitude limit exactly equals the predicted L$_{*}$(LBG) H magnitude; observations to the left would all be able to detect an L$_{*}$(LBG) galaxy at greater than 5-$\sigma$, while those to the right could only see galaxies brighter than L$_{*}$(LBG). The majority of the observations are sensitive to galaxies of L$_{*}$(LBG) or brighter, with only the two z$>$3 galaxies clearly beyond the limit of L$_{*}$(LBG) detectability. The 5-$\sigma$ magnitude limit is for the relatively compact object expected at high redshifts, FWHM = 0.4$\arcsec$, the median FWHM of our dim (H$>$20) galaxy detections. Larger FWHMs would shift the 5-$\sigma$ limit to brighter magnitudes, pushing down the points in the plot, making more of the observations insensitive to L$_{*}$(LBG) galaxies. 

Unfortunately, several effects decrease our actual sensitivity to the galaxies. The first important one is the inconsistent nature of the noise across the detector, particularly a roughly circular area of greater noise centered near but to the left of the coronagraphic hole, with a radius ranging from 0.75 to 2 pixels in size. This noise affected about 60\% of the coronagraph observations, but was only a minor effect, reducing the magnitude limit detectable in these areas by \ab 0.3 magnitudes. More important were the areas near the quasar that fell off the detector entirely, making it impossible to detect a DLA candidate there. This is critically important in three cases, BRI1500+0824, Q1610+2806, \& $[$HB89$]$0836+113, where the target acquisition program failed to find the quasar and instead of moving the quasar under the coronagraphic hole, moved it completely off the detector. These observations were still usable, however, because software recorded exactly how far off the detector the quasar was moved: 1.1, 0.4, and 2.3 arcseconds off the chip respectively. This means we will not find the DLA objects if they lie close to the quasar or on the unobserved side. Even so, these observations sample a good deal of area near the quasar that, when combined with the rest of the data, has statistical significance. We just can not say much about non-detections for those three sources. 

In most observations the quasar or coronagraphic hole was \ab 3.5 arcseconds from the edge of the detector, creating a region not far from the quasar that could completely hide its DLA candidate. Taking this unsurveyed area into account, we calculate that, if we exclude the three observations with the quasar off the edge, we observed 100\% of the area within 15$h^{-1}$ kpc of the quasar line of sight, 96\% of the area from 15 to 30$h^{-1}$ kpc, 75\% of the area from 30 to 45$h^{-1}$ kpc , and 54\% of the area from 45 to 60$h^{-1}$ kpc. This coverage trend continues downward from there, where we observe less than 7\% of the area at distances of 135 to 150$h^{-1}$ kpc from the quasar line of sight, which is roughly the limiting distance an object could lie from the quasar and still remain on the chip. There is, of course, also the area of the hole itself, a region with radius \ab 0.4\arcsec , that could hide the intervening DLA object. The same consideration also applies to the direct observations, where the area still unusable after PSF subtraction was very similar in size to the coronagraphic hole. This missing area at the quasar makes up approximately 7\% of the area within 15$h^{-1}$ kpc, meaning we actually observe only 93\% of the area within that distance. 

Finally, in the three cases of bright background quasars (JVAS1757+7539, H=15.2; $[$HB89$]$1157+014, H=15.7; $[$HB89$]$0454+039, H=15.4) the subtraction of leftover light from the image left large residuals that affected our ability to search within one arcsecond of the quasar. To test how severe this effect might be, in software we placed multiple sample galaxies of different known magnitudes all around the coronagraphic holes in these three images. For these three brightest quasars (QSOs), we found that the magnitude limit was only 0.3-0.7 magnitudes lower when the galaxy was placed \ab 0.75\arcsec \ (typically \ab 6$h^{-1}$kpc ) from the quasars. There was a low instance of ``catastrophic'' failures, where no galaxies of reasonable brightness could be recovered if they were placed at precisely the wrong spot (less than 1 in 10). Galaxies placed farther out than an arcsecond showed the same recovery rate as elsewhere within the same image. 

\section{Results}

Out of the 7000 square arcseconds, or 2 square arcminutes, imaged near \dla \ quasars, 31 galaxies were bright enough to be 5$\sigma$ or greater detections. If we remove those galaxies found near $[$HB89$]$0809+4822, the known DLA galaxy with $z_{abs}=0.437$ and previously identified optical counterparts, then the number of detected galaxies drops to 23. Of these 23 detected galaxies, we wish to determine which could be possible candidates for the absorbing DLA galaxies.

\subsection{DLA Candidate Criteria}

We first eliminate all galaxies that are simply too far away from the quasar line of sight to be possible absorber candidates. This means we need to know the expected impact paramater for DLAs from the quasar line of sight. Since few DLA hosts have been discovered, this number must be based on models of DLA HI distribution. The impact parameter will be small for compact, spherical objects and large for massive, extended disks. Since we seek to include all possible candidates, we will make the assumption that DLAs are the disk progenitors of today's spiral galaxies, which gives us the largest impact parameters. This may well be incorrect, but should provide us with an outer limit inside which more compact HI distributions would be expected to lie. In contrast, the spherical halo of clouds from the models of \cite{mcd99} predict typical impact parameters of just a few kpc, with the majority of systems lying within \ab 10 kpc.

We will start with the formula for the expected mean number density of DLAs for each z, $\left< n(z) \right>$, from \cite{lan91}:

\begin{eqnarray}
\left< n(z) \right> \approx 1.9 \times 10^{-2} \left( \frac {\xi}{1.5} \right)^2 \left( \frac {R_{*}}{11.5h^{-1}kpc} \right)^2 \nonumber \\
\times \left( \frac {f_{S}}{0.7} \right) \left( \frac {\Phi _{*}}{1.2 \times 10^{-2}h^{3}Mpc^{-3}} \right) \nonumber \\
\times \frac {\Gamma (1 + 2t -s)}{\Gamma (0.55)} (1 + \left< z \right>)(1 + 2q_{o}\left< z \right>)^{-1/2} \\ \nonumber
\end{eqnarray}

\noindent 
where $\xi$ is ratio of the HI limiting radius to the optical Holmberg Radius, R$_{*}$, $f_S$ is the fraction of galaxies that are spirals, $\Phi _{*}$ is the normalization for the luminosity function, $\Gamma $ is the gamma function, $t$ is the power-law index of the correlation between optical radius and luminosity, and s is the power-law index of the luminosity function. When \cite{lan91} input the fiducial choice of paramters ($\xi$=1.5, R$_{*}$=$11.5h^{-1}kpc$, t=0.4, $f_S$=0.7, $\Phi _{*}$=1.2 $\times 10^{-2}h^{3}Mpc^{-3}$, and $s$=1.25), they predicted a mean DLA number density at least a factor of 2 lower than what they were observing in their quasar spectra. This told them that either their fiducial values were wrong, or there has been significant evolution in the size or number density of galactic disks. We will consider the case where the effect is entirely due to an increase in galactic disk size, since this should give us the largest expected distances our candidate galaxies could be from the quasar line of sight. We will assume our L$_{*}$(LBG) galaxies obey the these same fiducial parameters, changing only in size, as we discuss below.

The radius of the typical HI disk that would produce a \dla \ line (N$_{H} >$ 2 $\times$ 10$^{20}$) is $\xi$R$_{*}$, which equals 17.25$h^{-1}$ kpc for the \cite{lan91} fiducial model. Looking at equation (1), we see that the mean number density observed is proportional to ($\xi$R$_{*})^{2}$, so we would only need to increase the HI disk radius by $\sqrt{2}$ to remove the factor of 2 discrepancy between predicted number density and that which is observed. However, this number density discrepancy is a lower limit and the discrepancy might be as high as a factor of 3-4 \citep{lan91,wol95}. This also assumes a $q_{o}$=0 universe. For a $q_{o}$=0.5 universe the number density discrepancy is much larger, as high as a factor of 6-8 \citep{lan91}. For this paper we will assume a $q_{o}$=0.1, which gives essentially the same results as the $q_{o}$=0 universe. This gives us an upper limit of 4 for the observed number density discrepancy. This means to match observed number densities to predicted densities using size evolution alone, we must double the typical $\xi$R$_{*}$ to 34.5$h^{-1}$ kpc. This is where we set the maximum radius for which we believe a L$_{*}$(LBG) galaxy could possibly still intercept the quasar light, producing the DLA. For the less extreme number density discrepancy factor of two, the maximum radius for a L$_{*}$(LBG) galaxy is only 25.1$h^{-1}$ kpc. These impact parameter limits, one showing no size evolution and the other a doubling in size, are plotted in Figure 6, scaled as (L/L$_{*}$(LBG))$^{0.4}$ to account for changes in predicted size with luminosity. It is important to recall that these lines represent the edge of the HI disk thick enough to produce a DLA. The {\it typical impact parameter} would lie to the left of the line, almost one third further in. It should be noted that local galaxies do show some spread in their fiducial values, $\xi$ and R$_{*}$, with the product of them changing roughly $\pm$50\% \citep{bos81,pet79}, so slightly larger values for the impact parameter are also possible in a few cases.  

In addition to removing candidates that lie too far from the line of sight, we can also eliminate those that are simply too bright to be galaxies at these high redshifts. For this we used the predicted L$_{*}$(LBG) galaxy magnitudes we calculated. If high-redshift galaxies continue to follow a Schecter luminosity function (s=1.25), as they seem to for z=3 and z=4 \citep{ste99}, it is unlikely that any of the DLA systems would be much brighter than one magnitude above an L$_{*}$(LBG) galaxy. This is despite the fact that brighter galaxies are predicted to have larger cross-sections for absorption. If one assumes the radius of a DLA system acts like today's galaxies, with size increasing as L$^{0.4}$ \citep{pet79}, then the cross section of our potential DLA galaxies should go as L$^{0.8}$. Weighting our Schechter function by this cross section and integrating we can determine the percentage of DLAs expected above any fraction of L$_{*}$(LBG). For a magnitude brighter than L$_{*}$(LBG), the percentage of expected DLAs is less than 3\%. Therefore we place our cut-off at one magnitude brighter than L$_{*}$(LBG), and only objects fainter than that are considered real DLA candidates. In Figure 6 we plot all candidates with their projected distances from the quasar line of sight versus their magnitude difference from a predicted L$_{*}$(LBG) H magnitude at the absorption redshift. (Graph actually shows 22 galaxies; the 23rd galaxy is associated with DLA \# 13 and is so bright it lies below the graph). Also plotted are two different boundary lines, depending on how much size evolution we allow, inside which we believe real DLA candidate systems would lie. These boundary lines are curved because of the assumed L$^{0.4}$ dependence of the radius.
 
Before counting possible DLA candidates, we should take a closer look at the strength of the DLA absorptions. For extremely large column densities of hydrogen it is likely that we are looking closer to the center than our previous $\xi$R$_{*}$ limit. If we assume that these galaxies have exponential disks with scale lengths similar to what is seen locally, \ab 4$h^{-1}$ kpc \citep{bos81}, but which would scale directly with their possible larger size (up to 8$h^{-1}$ kpc), we can predict how much closer to quasar we would expect to find them. Considering the uncertainties involved with this calculation, we did not remove any candidates that appeared to lie just a few $h^{-1}$ kpc beyond this new limit, choosing instead to concentrate on the objects with the highest column densities in our sample (log n(HI) = 21.4, 21.8, \& 21.4), all more than a factor of 10 (more than 2 scale lengths) above the \dla \ cut-off. These systems, $[$HB89$]$0000-263, $[$HB89$]$1157+014, and $[$HB89$]$1209+093 (DLAs \# 1, 10, \& 11), do not show any galaxies within the impact parameter limits on the graph, but do have a couple of candidate galaxies that are just beyond those limits. Their high column densities should further rule them out. These galaxies are marked with stars on Figure 6, and none should be considered plausible DLA candidates.

This leaves us with 3 of the 23 observed galaxies lying within the largest region, which assumes a doubling in size for high-redshift disks. Examining these candidates further, another problem appears. Two of the three candidates are about a half magnitude or brighter than an L$_{*}$(LBG) galaxy and at a large projected distance from the line of sight (\ab 30-40$h^{-1}$ kpc). Even accounting for the fact that we may be missing some of the dim DLA systems, these galaxies make up much too large a percentage of the sample. DLA systems between 0.5 and 1.5 L$_{*}$(LBG) should be found at three times the rate of those with magnitudes greater than 1.5 L$_{*}$(LBG). If both of these galaxies were identified with the DLA object, we get the unlikely result that 67\% of our candidates would be from this bright end of the luminosity function. We also point out that 10-12 of the discovered galaxies, half of those discovered, are located near this spot in the diagram. It is most likely that these galaxies are lower redshift galaxies, found around the average distance (\ab 10$\arcsec$ or 55h$^{-1}$ kpc for z=2) that one would measure from the coronagraphic hole for {\it randomly placed foreground field galaxies}. Our inability to distinguish these two detections from random field galaxies leads us to reject both galaxies as strong DLA candidates. The same dilemma applies to the two detections that were marginally too bright for our criteria (those for \#'s 8 \& 17$/$18). Even if one extended the cut-off to slightly brighter magnitudes, these two galaxies would still remain implausible DLA candidates.      
    
So in the end we found only one galaxy we consider to be a possible DLA system candidate, that associated with LBQS0010-0012 (\# 2) at a projected separation of 18.4 $h^{-1}$ kpc. If we assume no size evolution then there are no DLA candidates remaining, although the one associated with LBQS0010-0012 is near the limit and would only require small adjustments in impact parameter or L$_{*}$(LBG) magnitude to bring it over. It is important to note that we are being extremely generous with the allowed impact parameter from the quasar. For most of the high redshift DLA candidates suggested so far, the median impact parameter from the quasar is only \ab 8$h^{-1}$ kpc , with the larger impact parameters (i.e., quasar 2231+131) around 15$h^{-1}$ kpc \citep{mol98}. Most recently, \cite{kul00} and \cite{kul01} have claimed candidates for DLAs at z=1.89 toward LBQS 1210+1731 and at z=1.86 toward QSO 1244+3443, all with impact parameters around 1$h^{-1}$ kpc (0.25 arcseconds). They used NICMOS deep observations (2500-5000 seconds), each 3-6 times longer than our NICMOS snapshots. While many of these candidates may turn out to be unassociated with the DLA, or possibly not even real detections, every one of them is closer than our best candidate.   

\subsection{Search for Possible Overdensities}
 
We searched for any overdensity of galaxies near the quasar, but did not find any. We display a simple plot of number density versus distance from the quasar in Figure 7, which shows this lack of a rising trend toward the center. The number density observed (3-8$\times$10$^4$/deg$^2$ for H\ab 21.5-22.5), is in good agreement with other deep galaxy counts also taken with the F160W filter on NICMOS \citep{yan98}, so we are not finding galaxies at a higher rate in our quasar fields.

Six of the nineteen QSOs in our sample are also radio loud quasars: [HB89]0454+039 (PKS 0454+039), [HB89]0809+483 (3C 196), [HB89]1157+014 (PKS 1157+014), [HB89]1215+333 (MG2 J121728+3306), JVAS1757+7539, \& LBQS2359-0216B (NVSS J000150-015936). If we exclude the known overabundance of galaxies found near the intermediate-redshift quasar, [HB89]0809+483 (z$_{em}=0.871$), we find no overabundance of galaxies in these fields (4 galaxies in 5 fields) compared to the rest of the quasar sample (19 galaxies in 14 fields). Neither do we see any significant difference in the distributions of impact parameter or quasar brightness, although the sample is small. 

Finally, we added up all the flux from the discovered galaxies within different distances from the quasar line of sight and divided that by the number of observations. Excluding the quasars that actually lie off of the chip, we found an average H magnitude in observed galaxies of 21.8 for less than 30$h^{-1}$ kpc and 21.2 for less than 60$h^{-1}$ kpc. There are only two galaxies within 30$h^{-1}$ kpc, making it particularly vulnerable to small number statistics, but that is essentially the value one would expect to find in a random circle on the sky of the same size using NICMOS number counts (Yan et al. 1998) for galaxies between H=22.5 and H=19.0. There are seven galaxies within 60$h^{-1}$ kpc, which is also a relatively small number, but their flux actually lies below the predicted value (H = \ab 20.1). {\it Once again, we find no statisical evidence that our detected galaxies are associated with either the quasar or its \dla \ absorber.}

\subsection{Lack of L$_{*}$ Galaxies}

Our only high-redshift DLA candidate galaxy is that for LBQS0010-0012 (\# 2) with H=21.2 or \ab 0.65 L$_{*}$(LBG). No L$_{*}$(LBG) or greater potential candidates were see. From integrating a cross-section weighted luminosity function, we expected L$_{*}$(LBG) or brighter galaxies to be 17\% of our sample. If we eliminate our low-redshift candidates (\# 7) and the three acquisition failures (\#'s 8, 14, \& 15), we searched 18 DLAs, for which we should have found 3 candidates brighter than L$_{*}$(LBG). In fact, if we account for the sensitivities of all 18 observations, which could usually have detected  \ab 0.7 L$_{*}$(LBG) or better, we should have seen 5-6 DLA candidates total. Instead we saw only the one. 

It is not possible to obtain consistency with this measurement and both the $z=0$ and $z=3$ galaxy populations by varying the evolution.
One can choose different exponentially decaying star formation models that predict dimmer galaxies at z\ab 2, to try explaining the lack of DLA emission detections. For instance, a rapidly evolving model (short timescale) would fade rapidly as one went forward in redshift, making LBGs much less bright at $z$\ab 2. Alternatively, a slowly evolving model (long timescale) would brighten very slowly as one went back in redshift from $z=0$, making today's galaxies still quite dim at our DLA redshifts. The problem is finding a model that satisfies both these extreme conditions AND allows LBGs to become today's galaxies.

This situation is displayed in Figure 8, where we plot H absolute magnitude vs redshift. All points are measured data, where our measured upper limits are triangles and our detections are stars. L$_{*}$ for today's galaxies and that for Lyman-break galaxies are given as diamonds with error bars. The lines represent three different exponentially decaying star formation rate \citep{bru96} models, with timescales of 1 Gyr (dotted line), 4 Gyr (solid line), and 7 Gyr (dashed). They show both changes because of K-correction (bandpass shifting) and changes from evolution. The top figure shows the models fit to L$_{*}$(LBG) at $z=3.04$, while the bottom figure shows the same models fit to L$_{*}$ of today's ($z=0.0$) galaxies. Our preferred model is the the 4 Gyr solid line, as it evolves the $z=3$ luminosity function L$_{*}$ into the present day L$_{*}$.

If DLAs are taken from the same population as either today's galaxies or LBGs at $z=3$, then we would have expected to see three detections above the model line. For our 4 Gyr timescale moderate evolution, the majority of our upper limits lie below the line, a clearly inconsistent result. Models that dip below many of the upper limits do not allow a match between LBGs and today's galaxies. The rapid evolution (1 Gyr timescale) forward from $z=3$ LBGs would create galaxies more than a magnitude less luminous than today's galaxies, while the slow evolution (7 Gyr timescale) backward from today would imply galaxies 0.5-1.0 magnitudes too dim to be the detected LBGs. These evolutionary models are not the only models available, and one could conceivably add instantaneous bursts at just the right redshifts to match both low and high redshift points and explain the non-detections, but this sort of perfect conspiracy is unlikely in a large sample of objects.

At the redshifts of our DLAs, we are not seeing evolved L$_{*}$ galaxies, whether they be Lyman-break galaxies or those of the present day. Standard evolutionary models do not provide a lot of flexibility to explain this, if LBGs are to become today's galaxies. Either DLAs do not come from the same distribution as LBGs and present day galaxies, LBGs do not become present day galaxies, or both.           

After we submitted this paper, the work of \cite{war01} 
came to our attention. They surveyed a slightly smaller number of DLA systems
(15), but to greater depth (7 times longer exposures). Two of our quasars 
overlapped, [HB89] 1215+333 and LBQS 2206-1958. They did not find any 
DLA candidates in these 
fields either, at least down to the magnitudes to which our survey was 
sensitive. We analyzed their bright (and therefore comparable) detections in
the same way as those in our survey, finding 3-4 candidates
with reasonable separations and L$_{*}$(LBG) or greater. However, 
two of these DLAs
are at the same redshift as the background quasar. Thus the two objects 
detected may not be physically independent from the quasar.  
Disregarding this concern, their detection rate of
3 bright candidates out of 15 DLA fields would be roughly consistent 
with the hypothesis that DLA systems are normal galaxies, evolving from 
Lyman break systems into present day galaxies. However, it is also
consistent with our finding that DLA systems are significantly fainter than 
normal galaxies, based on only 1 candidate out of 18 fields.
 This inconsistency may have resulted from the small number statistics of 
both studies.

\section{Discussion}

We are not seeing galaxies near quasar lines of sight that contain \dla \ absorption. This result is very similar to the negative results of groups surveying DLA systems searching for the emission lines of Lyman \a \ \citep{low95} and H\a \ \citep{mal95,bun99}, suggesting that that their difficulties may not have been because the DLA systems do not produce strong emission lines, but because they do not produce much light at all. Our ability to look much closer to the quasar, 0.5 arcseconds or better, also weakens the argument that DLAs are not being found because they lie at such small impact parameters. In fact, the candidate seen by \cite{ara96} near Q1215+333 at a distance of 1.3 arcseconds is not seen in our data. With a K magnitude of 20.1, we should have picked it up, even with a red color (H-K $\geq$ 1.5). We tried models creating all stars in a single burst around z=10, but none were redder than H-K = 1.1 at z=2 \citep{bru96}, with most of our exponential models predicting H-K = 0.8-0.9. Colors redder than H-K = 1.2 required redshifts of z=3-3.5 and produced correspondingly dimmer galaxies. This makes it likely the former detection was a result of the very difficult subtraction of a bright PSF so near the quasar. It does not seem likely that we can continue to hide the majority of DLA systems by pushing them all into the tiny area (less than 0.8 arcsec$^2$ for our cornagraphic images) directly in front of the quasar, especially considering the extreme number evolution of galaxies that would require. 

Also not seen are the \cite{ste92} candidate for [HB89] 0000-263 (DLA \# 1), at a distance of 2.8$\arcsec$, and the \cite{ste95} and \cite{leb97} candidate for [HB89] 0454+039 (DLA \# 6), measured at two different distances from the quasar, 2.1$\arcsec$ and 0.8$\arcsec$ respectively. These last two may have been missed if the candidate were blue enough, R-H $\leq$ 3.4 for the [HB89] 0000-263 candidate and R-H $\leq$ 2.3 for [HB89] 0454+039. Models of high-redshift colors \citep{bru96} suggest that the former is likely (predicted R-H = 1.3-1.5 for exponential star formation models at z=3.4), while the latter is possible, but difficult to fit (predicted R-H = 2.3-3.5 for exponential star formation models at z=0.86). Galaxies as blue as R-H = 2.3 at z=0.86 require either strong continuous star formation or a strong starburst just prior to observation. 

All this leads us to speculate as to why we are not detecting light from almost any of these DLA systems. One possibility is that they not producing much light because they have not started making many stars by the period at which we are observing them. Most of our DLA sample comes from around z$\sim$2, so if most star formation occurred after that we would not be able to see it. One possible problem with this scenario is the previously observed decrease in comoving mass density of neutral gas seen in DLAs from z$\sim$3.5 to z=2 and then down into the present day. If DLAs are spiral galaxy progenitors, this is explained as the transformation of their neutral gas into stars. Initial work by \cite{wol95} indicated that as much one half of the neutral gas was depleted before z=2, meaning half of all stars should have been created before then and would therefore likely be visible to our survey. However, further work by \cite{smi96}, using some of the highest redshift DLAs known, showed a less dramatic neutral gas evolution, with only 20\% of the neutral gas processed by z=2. Also, as noted in the introduction, \cite{rao00} puts the whole idea of the decline of neutral gas in DLAs with time into doubt, showing no decline down into low redshifts. Neutral gas evolution aside, if DLAs are to become today's galaxies they must form stars at some point and a large amount of star formation is already measured to be underway before z=2 \citep{mad98}. Waiting until low redshifts to produce the vast majority of a galaxy's stars would create a problem with the number of highly luminous galaxies required at z$<$2.

Dust could also play a factor, absorbing light even at rest wavelength optical wavelengths. However there is no evidence for such large quantities of dust seen in the quasar spectra \citep{pei91,pet97}, requiring a most unlikely conspiracy of much higher dust quantities surrounding the star formation regions, but not along the quasar line of sight. 

Another possibility is that the DLA systems are not compact, high surface brightness objects like the spiral galaxies we study, but are instead diffuse, low-surface-brightness galaxies as suggested by \cite{jim99}. These galaxies would be inefficient at transforming their gas into stars, so would take longer to start forming stars, and would be harder to detect, even if their total magnitude equalled our predictions. If this is the case, the majority of DLAs are not like the LBGs discovered so far, and either these low surface brightness galaxies have to evolve into the high surface brightness spirals we see around us today, or these DLA systems will never become today's L$_{*}$ galaxies. This possibilty -- that DLAs are not high-redshift versions of present day disk galaxies -- could by itself explain all of our upper limits. They might be protogalactic blobs that will eventually fall into galaxies, helping them form or grow, or they might never become part of standard galaxies at all, being something entirely separate. 

While there may be a handful of DLA systems that we can detect in emission, the vast majority do not seem to be taken from the distributions of either $z=0$ or $z=3$ galaxies. Either they are not producing as many stars, something is absorbing their light, or their surface brightness is unusually low. From the evidence gathered so far, it appears that the distribution of DLA systems is inconsistent with both the evolution of Lyman-break galaxies forward in time and present day galaxies backwards. 

\acknowledgements

We would like to acknowledge and thank Lisa Storrie-Lombardi for providing her unpublished compilation of DLA systems and Wayne Webb for all his help in the early stages of the project. We would also like to acknowledge the assistance of NASA grant GO7329. 

\appendix

\section{Individual DLA Notes}

[HB89] 0000-263 --- \cite{ste92} identified an {\it R}=24.8 DLA candidate 2.8$\arcsec$ from the quasar's center to the southwest, evident only after PSF subtraction of the QSO image from their {\it G} and {\it R} frames. Subsequently, at least one Ly$\alpha$ emitting galaxy has been found in the field \citep{mac93}, although 87$\arcsec$ from the quasar, and as many as 14 other galaxies with similar broad band colors \citep{ste93}. We see no evidence for the \cite{ste92} object, but the area visible to the southwest of the quasar only extends \ab 3$\arcsec$ before reaching the edge of our detector. If the object lay slightly further to the southwest than measured by \cite{ste92}, it would have escaped our detection. However, with the limit R-H $\leq$ 3.4, modeling suggests it is unlikely we could have observed it either way (predicted R-H = 1.3-3.3).

LBQS 0010-0012 --- This is the only quasar in our survey for which we believe we have discovered a possible DLA candidate, lying 3.2$\arcsec$ to the southwest. If responsible for the \dla \ seen in the quasar spectrum, this H=21.2 object lies at a $z$=2.03, putting it at projected distance of 18.4 h$^{-1}$ kpc with a luminosity of \ab 0.65 L$_{*}$(LBG). 

LBQS 0013-0029 --- \cite{ara94} identified two K \ab 20 objects at distances of 4$\arcsec$ and 5$\arcsec$, to the south and west respectively. Unfortunately both reported positions lie off the edge of the detector. \cite{ge97} discovered molecular hydrogren absorption at the redshift of the DLA with a column density of N(H$_{2}$)=6.9(\p 1.6) $\times$ 10$^{19}$ cm$^{-2}$. 

[HB89] 0454+039 --- \cite{ste95} identified a possible candidate R=24.6 galaxy, 2.1$\arcsec$ to the north of the quasar, visible after subtracting the QSO light profile. \cite{leb97} took HST observations of this quasar, also finding an object to the northeast after subtraction of the quasar PSF, but at a closer distance (0.8$\arcsec$). We see no evidence for this object at either position. With our detection limit of H=22.3, the undetected galaxy would have to be fairly blue (R-H$\leq$2.3) for a z=0.86 galaxy.

[HB89] 0809+483 --- Also known as 3C 196, this object has a DLA at the moderate redshift of $z_{abs}$=0.437. This strongly damped object is also one of the 21 cm absorption systems \citep{fol88}. HST observations \citep{coh96,leb97}, reveal a barred spiral overlapping the quasar line of sight and at least one other candidate only an arcsecond away, which would be roughly 4h$^{-1}$ kpc away at the redshift of the absorber. We confirm both candidates, giving H magnitudes of 17.5 for the spiral and 19.6 for the other nearby companion. An L$_{*}$ galaxy at the absorber redshift is predicted to be \ab 16.9 mags. Also in our field are at least six other galaxies, with distances from the quasar and H magnitudes of 7.5$\arcsec$ and 20.1 mags, 7.5$\arcsec$ and 18.5 mags, 11.7 $\arcsec$ and 21.1 mags, 12.9$\arcsec$ and 20.7 mags, 14.0$\arcsec$ and 19.9 mags, and 14.2$\arcsec$ and 19.8 mags, of which only the faintest had not been previously seen. Using the F702W HST magnitudes from \cite{leb97} and their suggestion for conversion to the UBVRI magnitude system (R - {\it m}$_{702}$=0.35), we get the following R-H colors, given in the same order in which the galaxies were listed above: 2.81 (the spiral), 3.05 (other close companion), 3.74, 3.89, $\geq$4.4, 3.56, 4.4, and 3.76.  Models suggests that while the spiral and the other close companion have colors consisent with with a z=0.437 object (R-H = 2.3-3), the objects with R-H around four would be more consistent with the redshift of the quasar (z=0.871). This indicates a possible z=0.87 galaxy cluster at the QSO.  

[HB89] 0836+113 --- \cite{wol92} identified a Ly$\alpha$ emitting candidate at the redshift of the DLA ($z_{abs}$=2.466), a spatially resolved ``wispy'' object about 3-4$\arcsec$ to the north of the object. Work by \cite{low95} found no evidence of Ly$\alpha$ emission, but instead identified the previously seen emission as [OII]$\lambda$3727 from a z=0.788 Mg II absorber known to also exist along the quasar line of sight. \cite{ben97} discovered another object (K \ab 18.9, R-K = 4.7) 10.6$\arcsec$ to the south of the QSO, which they identify with a CIV system at z=1.82. There is also an object (K \ab 20.35) 7$\arcsec$ to the southwest, pointed out by \cite{wol92}, with a relative excess of emission in a narrow band filter tuned to Ly$\alpha$ emission at the redshift of the DLA. \cite{ben97}, however, point out that it may be associated instead with a possible z=0.37 Mg II system. Unfortunately, this quasar was one of our acquisition failures, placing it 2$\arcsec$ off the detector to the northeast. We do see the \cite{ben97} object at a distance 10.4$\arcsec$, for which we measure an H of 19.3. This gives a color of H-K = 0.4, which is inconsistent with a z=1.82 object, for which modeling predicts something more like 0.8-1.1, and is hard to reconcile with the measured R-K color. We also believe we see the southwest object, however we place it at a distance of 8.8$\arcsec$ (not 7\arcsec ) with H=21.5. Its color (H-K =1.15) is consistent with z=0.37.   

[HB89] 1157+014 --- This quasar has long been known to possess a 21 centimeter absorption at $z_{abs}$=1.94 \citep{wol81}, the same as the DLA system.

[HB89] 1209+093 --- This object was included in the \cite{low95} negative search for Ly$\alpha$ emission at the redshift of DLA systems.

[HB89] 1215+333 --- \cite{ara96} identified a possible candidate K=20.1 galaxy, 1.3$\arcsec$ to the east of the QSO, visible after PSF subtraction of the quasar light. It was also included in the \cite{low95} search for Ly$\alpha$ emission at the redshift of DLA systems. Our observations find nothing at that position, giving a limit of H-K $\geq$1.5, too red for a galaxy at the redshift of the absorber (z=2). 

LBQS 2206-1958 --- \cite{ber92} imaged this field to identify known intervening Mg II systems. They successufully discovered three candidates, ranging in distance from 6 to 13.5$\arcsec$, and spectroscopically confirmed two of them (z=0.755, z=1.017). The third galaxy, at a distance of \ab 11$\arcsec$ to the northwest, is tentatively identified with a z=0.948 Mg II system, but they could not confirm it. Only the Mg II system at a distance of 6$\arcsec$ (r = 22.6) lies within the area observed by our observation, for which we find a magnitude of H=19.5. This gives us an approximate R-H $\approx$ 3.1, entirely consistent with a z=0.755 galaxy.  

[HB89] 2348-011 --- Also known as UM 184, this object was included in H$\alpha$ searches by both \cite{tep98} and \cite{bun99}. In neither instance were any H$\alpha$ candidates discovered.

\newpage

\figcaption[colbertj.Fig1.ps]{Observation of JVAS1757+7539 (DLA \# 16). Top is before scattered QSO light subtraction, showing the spilled light remaining even after the quasar is moved underneath the coronagraphic hole. Below is the same image after quasar light subtraction, with an arrow pointing to the exact spot of the coronagraphic hole. No real information lies within this hole and therefore anything within it should be ignored. Note that while most of the light has been removed in this way, some artifacts still remain, such as the tiny blob just above the circle, which is seen in repeatedly in many subtractions. Each image is 19.2$\arcsec$ on a side.}

\figcaption[colbertj.Fig2.ps]{Observation of LBQS0010-0012 (DLA \# 2). Top is before light subtraction, with both star (on left) and quasar (on right) directly imaged. Below is same image after star PSF has been scaled and subtracted from the quasar, with an arrow pointing to the position of the quasar. The object sitting to the southwest (up and to the left) of the quasar is the best DLA candidate we found in this survey.}
  
\figcaption[colbertj.Fig3a.ps,colbertj.Fig3b.ps]{These are light subtracted close-ups of all successful coronagraph observations, roughly 5$\arcsec \times$5$\arcsec$, varying slightly in an attempt to include all nearby objects of possible interest. The length of each arm of the compass is exactly 1.5$\arcsec$ (corresponding to \ab 8.5h$^{-1}$ kpc at z=2.5), providing an exact scale for each image. A circle marks the location of the coronagraphic hole, within which no real information exists. a: Starting at top left, reading across then down, the images are $[$HB89$]$0000-263 (DLA \# 1), LBQS0013-0029 (DLA \# 3), LBQS0102-0214 (DLA \# 4), $[$HB89$]$0249-2212 (DLA \# 5), $[$HB89$]$0454+039 (DLA \# 6), and $[$HB89$]$0809+4822  (DLA \# 7). b: Starting at top left, reading across then down, the images are $[$HB89$]$1157+0128 (DLA \# 10),  $[$HB89$]$1215+333 (DLA \# 12),  BRI1346-0322 (DLA \# 13),  JVAS1759+7539 (DLA \# 16), LBQS2206-1958 (DLA \# 17/18), and  $[$HB89$]$2348-011 (DLA \# 19/20). The observation of LBQS2359-0216B (DLA \# 21/22) is not included because of the severe noise effecting that image.}   

\figcaption[colbertj.Fig4.ps]{These are the PSF subtracted images of the three direct observations, roughly 5$\arcsec \times$5$\arcsec$, varying slightly in an attempt to include all nearby objects of possible interest. The length of each arm of the compass is exactly 1.5$\arcsec$, providing an exact scale for each image. A circle marks the location of the quasar. The top left is LBQS0010-0012 (DLA \# 2), the top right is $[$HB89$]$1151+068 (DLA \# 9), and the center bottom is $[$HB89$]$1209+093 (DLA \# 11).}

\figcaption[colbertj.Fig5.ps]{This figure represents our sensitivity to the detection of DLAs. Plotted are the 5$\sigma$ limits reached for all the target DLA quasar observations vs. the H magnitudes predicted for an L$_{*}$(LBG) galaxy at the redshift of the known DLA. Coronagraph observations are shown as diamonds, direct observations as stars, and acquisition failures as triangles. The solid line running through the figure shows where our limits would just allow us to see an L$_{*}$(LBG) galaxy or brighter, while the dashed line to the right indicates where we would only be able to see a galaxy that was one magnitude brighter than L$_{*}$(LBG) or brighter, and the dashed line to the left where we could see a galaxy down to one magnitude fainter than L$_{*}$(LBG). The dotted line shows cases in which we could see a galaxy as dim as 1/10 L$_{*}$(LBG). This means the farther to the left of the solid line an observation lies in this diagram, the dimmer a DLA object could be and still be discovered. $[$HB89$]$0809+4822 (DLA \# 7) is not included on this diagram, as its low relative redshift (z$_{abs}$=0.437) indicates a predicted L$_{*}$(LBG) magnitude (H=17.1) that places it well off the left side of the figure.} 

\figcaption[colbertj.Fig6.ps]{This figure plots 23 of the discovered galaxies (excluding those associated with DLA \# 7) by their distance from the quasar in projected h$^{-1}$ kpc, assuming the galaxy lies at the redshift of the DLA absorption, vs. their H magnitude difference from our predicted L$_{*}$(LBG) galaxy. The various lines demonstrate the limits inside which we believe it could be possible for the observed galaxies to be DLA systems. The inner dot-dashed line represents no size evolution for a disk galaxy, while the dashed line shows an increase in disk size by a factor of 2. Galaxies marked with stars indicate those associated with DLAs of extremely high column densities, effectively requiring these galaxies to be even closer to the quasar line of sight. Note the large number of galaxies located in the center of the diagram. Two of these barely satisfy the separation requirements, but are statistically unlikely. Only the object associated with LBQS0010-0012 (DLA \#2), at a projected distance of 18.4 h$^{-1}$ kpc, is considered a viable DLA candidate.}

\figcaption[colbertj.Fig7.ps]{The number density of observed galaxies, binned by their distance from the quasar. Horizontal bars are bin size, while vertical bars represent 1$\sigma$ error, assuming Poissonian statistics. There is no clear trend showing any rise or fall towards the quasar.}

\figcaption[colbertj.Fig8.ps]{a) Plot of H absolute magnitude vs redshift. Our coronagraph upper limits are empty triangles, the direct measurement upper limits are solid triangles, and our measured DLA candidate detections are stars. L$_{*}$ for today's galaxies and that for Lyman-break galaxies are given as diamonds with error bars. The lines represent three different exponentially decaying star formation rate BC96 models, with exponential timescales of 1 Gyr (dotted line), 4 Gyr (solid line), and 7 Gyr (dashed). The models are fit to L$_{*}$(LBG) at $z=3.04$. b) This plot is the same as 8a, only the models are fit to L$_{*}$ of today's ($z=0.02$) galaxies.}

\begin{deluxetable}{llccccccccc}
\tablecaption{\bf \Dla \ Targets} 
\tablehead{
\colhead{\#} & 
\colhead{Quasar} & 
\colhead{m$_V$} & 
\colhead{m$_H$} & 
\colhead{z$_{em}$} & 
\colhead{z$_{abs}$} & 
\colhead{log} & 
\colhead{L$_{*}$} & 
\colhead{5$\sigma$} &  
\colhead{Obs.} &
\colhead{REF} \\ 
& & & & & & 
\colhead{n(HI)} & 
\colhead{H Mag} &
\colhead{Limit} & 
\colhead{Type\tablenotemark{a}} &
}
\startdata 
1 &  $[$HB89$]$0000-263 & 18.0 & 15.8 & 4.10 & 3.40 & 21.4 & 22.7 & 21.4 & C & 1 \\
2 &  LBQS0010-0012 & 18.0 & 17.4 & 2.14 & 2.03 & 20.8 & 20.8 & 22.1 & D & 2 \\
3 &  LBQS0013-0029 & 17.0 & 16.0 & 2.08 & 1.97 & 20.1 & 20.7 & 22.0 & C & 2 \\
4 &  LBQS0102-0214 & 18.0 & 16.9 & 1.98 & 1.74 & 20.6 & 20.3 & 22.2 & C & 2 \\
5 &  $[$HB89$]$0249-222 & 18.4 & 16.2 & 3.20 & 2.83 & 20.2 & 21.8 & 22.2 & C & 2 \\
6 &  $[$HB89$]$0454+039 & 16.5 & 15.4 & 1.35 & 0.860 & 20.8 & 18.4 & 22.3 & C & 2 \\
7 &  $[$HB89$]$0809+483 & 17.8 & 16.3 & 0.871 & 0.437 & 20.2 & 16.9 & 20.9 & C & 2 \\
8 &  $[$HB89$]$0836+113 & 18.8 & 17.8 & 2.70 & 2.47 & 20.6 & 21.4 & 21.7 & A & 3 \\
9 &  $[$HB89$]$1151+068 & 18.8 & 16.6 & 2.76 & 1.77 & 21.3 & 20.4 & 22.0 & D & 3 \\
10 & $[$HB89$]$1157+014 & 17.7 & 15.7 & 1.99 & 1.94 & 21.8 & 20.7 & 22.3 & C & 2 \\
11 & $[$HB89$]$1209+093 & 18.5 & 17.4 & 3.30 & 2.58 & 21.4 & 21.5 & 22.0 & D & 1 \\
12 & $[$HB89$]$1215+333 & 17.5 & 17.1 & 2.61 & 1.99 & 20.95 & 20.7 & 21.6 & C & 3 \\
13 & BRI1346-0322 & 18.8\tablenotemark{b} & 17.0 & 3.99 & 3.73 & 20.3 & 23.2 & 22.3 & C & 4 \\
14 & BRI1500+0824 & 18.8\tablenotemark{b} & 17.6 & 3.94 & 2.80 & 20.4 & 21.7 & 22.2 & A & 4 \\
15 & Q1610+2806 & 19.2\tablenotemark{b} & 18.2 & 3.51 & 2.59 & 20.6 & 21.5 & 22.1 & A & 5 \\
16 & JVAS1757+7539 & 16.1\tablenotemark{b} & 15.2 & 3.05 & 2.63 & 20.8 & 21.5 & 22.3 & C & 6 \\
17 & LBQS2206-1958 & 17.3 & 15.7 & 2.56 & 1.92 & 20.7 & 20.6 & 21.0 & C & 7 \\
18 & LBQS2206-1958 & 17.3 & 15.7 & 2.56 & 2.08 & 20.4 & 20.9 & 21.0 & C & 7 \\
19 & $[$HB89$]$2348-011 & 18.0 & 17.0 & 3.01 & 2.43 & 20.5 & 21.3 & 21.5 & C & 3,7 \\
20 & $[$HB89$]$2348-011 & 18.0 & 17.0 & 3.01 & 2.61 & 21.3 & 21.5 & 21.5 & C & 3,7 \\
21 & LBQS2359-0216B & 18.0 & 17.3 & 2.82 & 2.10 & 20.9 & 20.9 & 20.9 & C & 1 \\
22 & LBQS2359-0216B & 18.0 & 17.3 & 2.82 & 2.16 & 20.3 & 21.0 & 20.9 & C & 1 
\enddata
\tablenotetext{a}{C: Coronagraph, D: Direct Image, A: Acquisition Failure}
\tablenotetext{b}{Magnitude given here is a R magnitude.}
\tablerefs{(1) Lanzetta et al. 1991; (2) Wolfe et al. 1995; (3) Wolfe et al. 1986; (4) Storrie-Lombardi et al. 1996; (5) L. J. Storrie-Lombardi 1999, private communication; (6) Prochaska \& Wolfe 1999; (7) Turnshek et al. 1989 }
\end{deluxetable}

\begin{deluxetable}{cccccc}
\tablecaption{\bf Complete List of Discovered Objects} 
\tablehead{
\colhead{Associated} & 
\colhead{z$_{abs}$} &
\colhead{Separation} & 
\colhead{Separation} & 
\colhead{H Mag} & 
\colhead{Mag Diff.} \\
\colhead{DLA \#} & & 
\colhead{(h$^{-1}$ kpc)} & 
\colhead{(arcsec)} & & 
\colhead{from L$_{*}$(LBG)\tablenotemark{a}}
}
\startdata
      1  &    3.40  &    53.3  &    9.65  &    20.7  &   -2.0  \\
      2  &    2.03  &    18.4  &    3.24  &    21.2  &   0.4 \\
      3  &    1.97  &    62.7  &    11.05 &     20.0 &   -0.7 \\
      4  &    1.74  &    36.4  &    6.47  &    21.9  &    1.6 \\
      4  &    1.74  &   46.4   &   8.24   &   19.4   &  -0.9 \\
      4  &    1.74  &    80.9  &    14.35 &     19.4 &    -0.9 \\
      6  &   0.860  &   76.1   &   15.51  &    17.4  &   -1.0 \\
      6  &   0.860  &    56.2  &    11.46 &     19.8 &    1.4 \\
      7  &   0.437  &   6.1    &  1.67    &  17.5    &  0.6 \\
      7  &   0.437  &    4.5   &   1.22   &   19.6   &   2.7 \\
      7  &   0.437  &    27.3  &    7.48  &    18.5  &    1.6 \\
      7  &   0.437  &   27.3   &   7.46   &  20.1    &  3.2 \\
      7  &   0.437  &   42.8   &   11.7   &   21.1  &    4.2 \\
      7  &   0.437  &    51.2  &    13.99 &    19.9 &     3.0 \\
      7  &   0.437  &    47.2  &    12.90 &    20.7 &     3.8 \\
      7  &   0.437  &    51.9  &    14.19 &    19.8 &     2.9 \\
      8  &   2.47   &    45.2  &    7.97  &   20.3  &  -1.1 \\
      8  &   2.47   &   59.3   &   10.46  &   19.3  &   -2.1 \\
      8  &   2.47   &   50.0   &   8.82   &  21.5   &    0.1 \\
      10 &    1.94  &    23.9  &    4.21  &   21.8  &     1.1 \\
      11 &    2.59  &    73.2  &    12.94 &    20.3 &    -1.2 \\
      11 &    2.59  &   63.6   &   11.24  &   20.5  &   -1.0 \\
      13 &    3.73  &    91.1  &    16.73 &    20.7 &    -2.5 \\
      13 &    3.73  &   53.5   &   9.82   &  19.1   &  -4.1 \\
      15 &    2.59  &   48.2   &   8.47   &  21.3   &   -0.2 \\
      15 &    2.59  &   103.4  &    18.27 &    22.2 &    0.7 \\
      16 &     2.63 &    32.6  &    5.77  &   21.2  &  -0.3 \\
      17/18 &  1.92 &    34.4  &    6.07  &   19.5  &   -1.1/-1.4 \\
      19/20 &  2.43 &    54.9  &    9.68  &   20.0  &   -1.3/-1.5 \\
      19/20 &  2.43 &    60.9  &    10.73 &    19.0 &    -2.3/-2.5 \\
      19/20 &  2.43 &    18.8  &    3.32  &   19.2  &   -2.1/-2.3 
\enddata
\tablenotetext{a}{Negative numbers indicate objects brighter (smaller magnitude) than a predicted L$_{*}$ galaxy.}
\end{deluxetable}
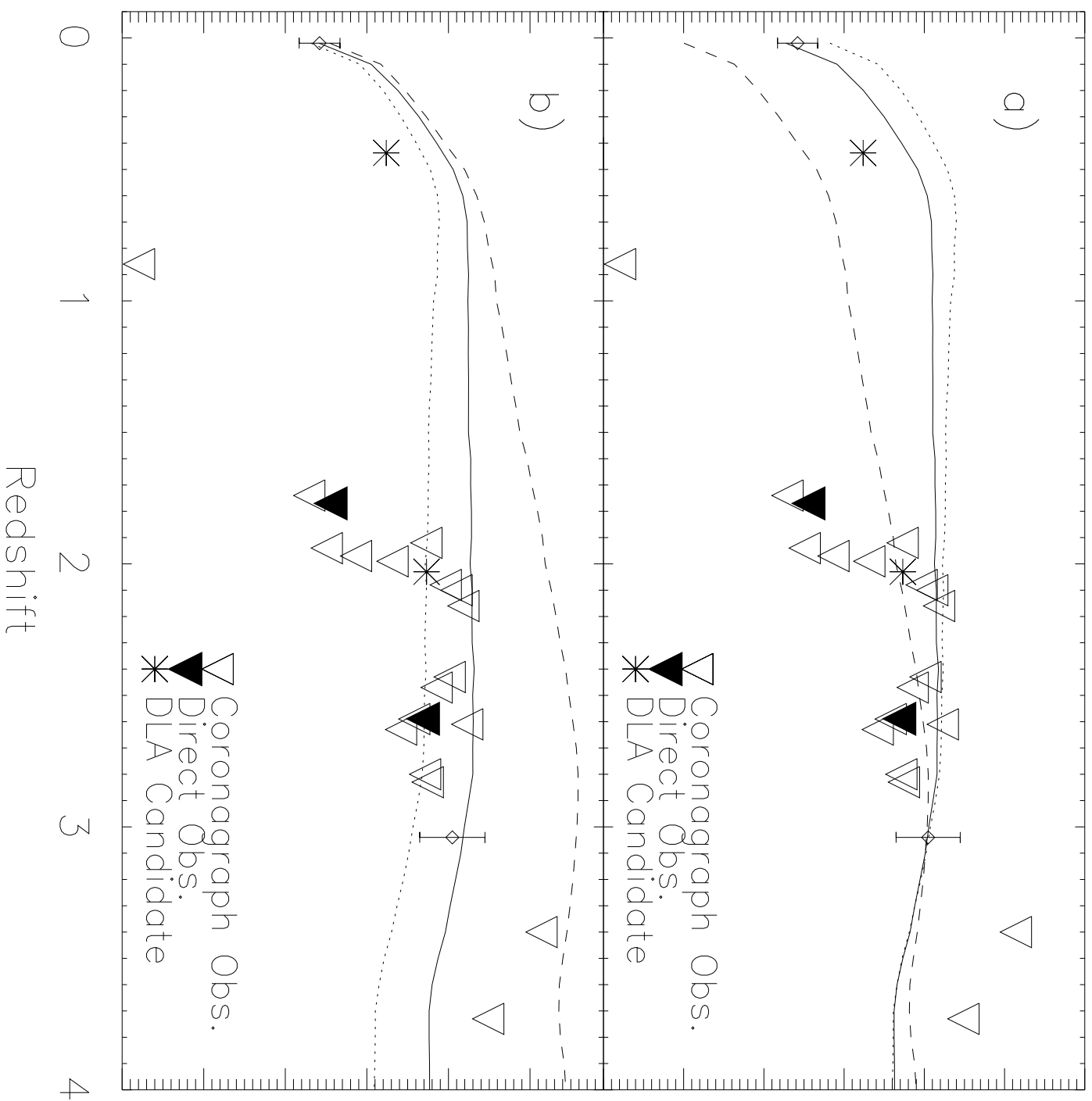
\end{document}